\documentclass[twoside,10pt]{article}
\usepackage{adjustbox}
\usepackage{extsizes}
\usepackage{silence}
\usepackage[super,sort&compress,comma]{natbib} 
\usepackage[version=3]{mhchem}
\usepackage[left=1.5cm, right=1.5cm, top=1.785cm, bottom=2.0cm]{geometry}
\usepackage{balance}
\usepackage{lmodern}
\usepackage{mathptmx}
\usepackage{multirow}
\usepackage{tcolorbox}
\usepackage{etoolbox}
\usepackage{sectsty}
\usepackage{graphicx} 
\usepackage{lastpage}
\usepackage[format=plain,justification=justified,singlelinecheck=false,font={stretch=1.125,small,sf},labelfont=bf,labelsep=space]{caption}
\usepackage{float}
\usepackage{fancyhdr}
\usepackage{fnpos}
\usepackage[english]{babel}
\addto{\captionsenglish}{%
  
}
\usepackage{array}
\usepackage{droidsans}
\usepackage{charter}
\usepackage[T1]{fontenc}
\usepackage[dvipsnames]{xcolor}
\usepackage{setspace}
\usepackage[compact]{titlesec}

\usepackage{epstopdf}
\definecolor{cream}{RGB}{222,217,201}

\newcommand{\Tp}{\hat T_{\rm p}}
\newcommand{\Te}{\hat T^{\rm ext}}
\newcommand{\Tt}{\hat T}
\newcommand{\Hn}{\hat{\mathcal{H}}_N^{(\rm CC)}}
\newcommand{\Hp}{\hat{\mathcal{H}}_N^{(\rm pCCD)}}

\newcommand{\Hfp}{\hat{\mathcal{H}}_N^{(\rm fpCC)}}
\newcommand{\Hfpl}{\hat{\mathcal{H}}_N^{(\rm fpLCC)}}
\usepackage{physics}
\usepackage{bm}
\usepackage{soul}
\usepackage[colorinlistoftodos]{todonotes}
\usepackage{hyperref}

\titleformat{\section}          
  {\bfseries\fontsize{12pt}{14pt}\selectfont} 
  {\thesection}                
  {1em}                        
  {}                           
\begin{document}

\begin{flushleft}
{\fontsize{18pt}{20pt}\selectfont\textbf{EOM-fpCCSD: An Accurate Alternative to EOM-CCSD for Doubly Excited and Charge-Transfer States.$\dag$}}
\end{flushleft}
{\fontsize{12pt}{14pt}\selectfont Katharina Boguslawski\textsuperscript{*a} and Paweł Tecmer\textsuperscript{b}}\\[2ex]
{\textit{Institute of Physics, Faculty of Physics, Astronomy and Informatics,\\ 
Nicolaus Copernicus University in Toruń, \\ 
Grudziądzka 5, 87-100 Toruń, Poland} }\\[4ex] 
*Email: k.boguslawski@umk.pl

\vspace{0.5cm}

We introduce a new equation-of-motion coupled-cluster method based on a pair coupled-cluster doubles (pCCD) reference, termed frozen-pair EOM-CCSD (EOM-fpCCSD).
This approach combines the computational efficiency of the pCCD ansatz with a dynamical correlation correction, enabling a reliable description of electronically excited states within the EOM framework.
The method has been implemented in the open-source PyBEST software package.
Its performance is systematically benchmarked against standard EOM-CCSD and its pair-tailored variant (EOM-ptCCSD), using both canonical Hartree–Fock and pCCD natural orbitals.
For charge-transfer (CT) excitations taken from the QUEST database, EOM-fpCCSD yields excitation energies very close to those of EOM-CCSD, outperforming EOM-ptCCSD, as well as to the theoretical best estimates (TBEs).
Working within the localized pCCD natural orbital basis allows us to determine the directed CT character, which quantifies the directed charge flow from one molecular domain to another. 
Numerical results show that EOM-fpCCSD, EOM-CCSD, and EOM-ptCCSD provide nearly identical descriptions of the directed CT character, despite changes in excitation energies.
The true advantage of EOM-fpCCSD becomes evident for the challenging QUEST subset of doubly excited states.
While EOM-ptCCSD performs similarly to standard EOM-CCSD, EOM-fpCCSD significantly outperforms both methods for these problematic states compared to TBEs.
In addition to improving the accuracy of excitation energies, EOM-fpCCSD also converges for several states that standard EOM-CCSD and EOM-ptCCSD fail to converge.
These results demonstrate that EOM-fpCCSD offers a promising and computationally efficient route toward a more accurate description of complex electronic excitations.
\vspace{1cm}
\section*{Introduction}
The equation-of-motion coupled-cluster singles and doubles method (EOM-CCSD)~\cite{eom-cc-rowe-rmp-1968, eom-cc-cpl-1989, eom-ccsd-bartlett-jcp-1993, bartlett-eom-cc-wires-2012,eom-cc-monika-book-chapter-2020} is one of the most widely used and reliable wave-function-based approaches for computing low-lying electronic excitation energies in atoms and molecules.
Thanks to its balanced treatment of dynamical correlation and its formal $\mathcal{O}(N^6)$ scaling, EOM-CCSD offers an excellent compromise between accuracy and computational cost.
Recent extensions include four-component relativistic formulations based on the Dirac–Coulomb Hamiltonian~\cite{4c-eom-ccsd-pra-2014, 4c-eom-ccsd-jcp-2018} as well as implementations for periodic systems.~\cite{eom-ccsd-for-solids-jctc-2017,eom-ccsd-performance-for-solids-jcp-2024}
Although more accurate variants incorporating, among others, triple excitations in both the cluster and excitation operators have been developed~\cite{eom-ccsdt-jcp-2001, active-space-eom-ccsdt-jcp-2001, cr-eom-ccsd_t-jcp-2004,cr-eom-ccsd-porhyrin-jcp-2010,cr-2-2-eom-piecuch-jctc-2012, eom-ccsd_t_a-jcp-2016} and alternative theories to mediate some of the drawbacks of conventional EOM-CCSD have been devised,~\cite{st-eom-cc-jcp-1995,multi-level-cc-jcp-2014,eom-vertical-ee--dutta-jctc-2018,mlccsd-core-excitations-jctc-2020, ascc-jpcl-2025, block-correlated-gvb-eom-jcpl-2025,eom-four-block-correlated-cc-jpcl-2026} it remains a popular choice for routine calculations.
It typically outperforms time-dependent density functional theory (TD-DFT) for low-lying excitations and can deliver accuracy comparable to high-level multireference methods such as MRCI~\cite{mrci-cr-2012} and CASPT2 unless the system is truly multireference.~\cite{casscf-roos-cp-1980, caspt2-jcp-1990, caspt2-jcp-1992, caspt2-pulay-perspective-ijqc-2011}
EOM-CCSD is generally considered one of the most reliable single-reference methods for valence and charge-transfer (CT) excitations.~\cite{eom-cc-benchmark-ct-states-jctc-2020}
Its most serious weakness is the poor description of states with significant doubly excited character, where errors can be large (up to several eVs), and the method may even fail to converge.~\cite{quest-jctc-2025}
To that end, the theoretical description of doubly-excited states remains an active field of research.~\cite{ascc-jctc-2024, ascc-jctc-2025}

Furthermore, the accurate quantum-chemical description of electronically excited states exhibiting pronounced charge transfer and/or double-excitation character remains a major challenge in photochemistry. Such states are central to key technological advances in organic solar cells,~\cite{ct-importance-osc-acs-appl-mat-2017, pccd-perspective-jpcl-2023, pccd-delaram-rsc-adv-2023, eom-pccd+s-ct-analysis-jctc-2025} singlet fission,~\cite{singlet-fission-cr-2010, doubly-excited-state-singlet-fission-chem-photo-chem-2021} thermally activated delayed fluorescence,~\cite{tadf-jpcl-2019, tadf-nature-2022} and organic light-emitting diodes in general.
To address this challenge, novel alternatives to the conventional EOM-CCSD framework are needed that preserve its favorable $\mathcal{O}(N^6)$ scaling while significantly improving the description of electronically excited states with substantial charge-transfer or doubly excited character.
Such developments are essential for advancing our understanding of organic electronic materials and for establishing reliable excited-state structure–property relationships.
In this work, we present a new family of EOM-CC models for electronically excited states, implemented in the open-source PyBEST software package.~\cite{pybest-paper-cpc-2021, pybest-paper-update1-cpc-2024}
These models are built upon the pair Coupled Cluster Doubles (pCCD) ansatz,~\cite{ap1rog-lcc-jctc-2015, oo-ap1rog-prb-2014, tamar-pccd-jcp-2014} known to perform well for electronic structures with quasi-degenerate states and strong electron correlation effects, allowing to model the bond-breaking process of electron-pair dominant bonds.~\cite{diatomics-oo-ap1rog-jpca_2014, ps2-ap1rog-jcp-2014,ap1rog-singlet-gs-actinides-pccp-2015,pccd-correlation-analysis-prb-2016}
A simple extension to electronic excited states within the EOM formalism has been introduced as the EOM-pCCD+S model.\cite{eom-pccd-jcp-2016, eom-pccd-erratum-jcp-2017}
Yet, to improve the description of electronic excited states, a dynamic energy correction is needed in the model.~\cite{eom-pccd-lccsd-jctc-2019, eom-pccd+s-ct-analysis-jctc-2025, optimal-excited-states-cc-reference-carter-fenk-jctc-2025, lr-pccd-jctc-2024}
One example is the linearized coupled-cluster correction on top of pCCD reference states,~\cite{ap1rog-lcc-jctc-2015, eom-lcc-jcp-2015, diagrammtic-lcc-carter-fenk-jpca-2025} which improves the electronic excited-state description.~\cite{eom-pccd-lccsd-jctc-2019, pccd-ee-f0-actinides-pccp-2019,pccd-static-embedding} 
In this work, we derive, implement, and benchmark a new EOM-pCCD-based approach: the frozen-pair EOM-CCSD method, denoted EOM-fpCCSD.
Its performance is systematically compared with the linearized version of the same ansatz (EOM-fpLCCSD) and the pair-tailored variant (EOM-ptCCSD).
We evaluate their performance against canonical Hartree–Fock orbitals for excited states with doubly excited character and charge-transfer states, taking advantage of variationally optimized pCCD natural orbitals for the latter.~\cite{oo-ap1rog-prb-2014, ap1rog-non-variational-orbital-optimizarion-jctc-2014, diatomics-oo-ap1rog-jpca_2014}
Our test set comprises dedicated charge-transfer and doubly excited subsets from the QUEST database.~\cite{quest-jctc-2025}
\section*{Methods}
In this work, we discuss several different excited-state extensions for a pCCD-based reference state.
Our starting point is the pCCD ground-state wavefunction,~\cite{limacher-ap1rog-jctc-2013, tamar-pccd-jcp-2014, pccp-geminal-review-pccp-2022}
\begin{equation}\label{eq:pccd}
\ket{\rm pCCD} = \exp \left (  \sum_{i=1}^{\rm occ} \sum_{a=1}^{\rm virt} t_{i\bar{i}}^{a\bar{a}} 
    \hat a_a^{\dagger}  \hat a_{\bar{a}}^{\dagger}\hat a_{\bar{i}} \hat a_{i}  \right ) \ket{\Phi_0} = e^{\Tp} \ket{\Phi_0},
\end{equation}
where $\Tp$ is the electron-pair excitation operator, containing electron creation ($\hat a^\dagger_p$ for $\alpha$ spin and $\hat a^\dagger_{\bar{p}}$ for $\beta$ spin) and annihilation ($\hat a_p$ for $\alpha$ spin and $\hat a_{\bar{p}}$ for $\beta$ spin) operators and the electron-pair cluster amplitudes $t_{i\bar{i}}^{a\bar{a}}$.
In the above equation, $| \Phi_0 \rangle$ is some reference determinant and the sum runs over all occupied $i$ and virtual $a$ orbitals of this reference state.
In the case of pCCD, this reference determinant (including its orbitals) is typically optimized and does not correspond to the canonical Hartree--Fock solution.~\cite{oo-ap1rog-prb-2014,tamar-pccd-jcp-2014,ap1rog-manual-orbital-rotations-mp-2014,ps2-ap1rog-jcp-2014,ap1rog-non-variational-orbital-optimizarion-jctc-2014}
Thus, by construction, pCCD describes electron correlation effects restricted to electron pairs, and the cluster operator spans only the seniority-zero sector of the wavefunction expansion.~\cite{limacher-ap1rog-jctc-2013,diatomics-oo-ap1rog-jpca_2014,tamar-pccd-jcp-2014, oo-ap1rog-prb-2014, ps2-ap1rog-jcp-2014, ap1rog-manual-orbital-rotations-mp-2014, seniority-zero-mean-field-wfn-jcp-2025}
Although this restriction turns out to be qualitatively and quantitatively acceptable to model strong electron correlation,~\cite{diatomics-oo-ap1rog-jpca_2014, ap1rog-singlet-gs-actinides-pccp-2015,pccd-correlation-analysis-prb-2016, state-specific-oopccd-jctc-2021}  correlation effects due to the missing broken-pair states, commonly attributed to be of dynamical nature, are crucial to reach chemical or spectroscopic accuracy.
The missing dynamical correlation can be accounted for using, for instance, perturbation theory~\cite{ap1rog-pt2b-pccp-2014,pccd-ptx-jctc-2017, seniority-0-canonical-transformation-theory-jcp-2026} or CC corrections.~\cite{frozen-pccd-jcp-2014,ap1rog-lcc-jctc-2015,pccd-tailored-jctc-2022}
In this work, we will focus on excited-state extensions of the latter corrections.

Essentially, CC corrections on top of pCCD can be understood in terms of tailored coupled cluster (tCC) theory,~\cite{tailored-cc-oliphant-jcp-1991,tailored-cc-implemntation-oliphant-jcp-1992,tailored-cc-jcp-1993,tailored-cc-jcp-1994} that imposes the pCCD wavefunction as the fixed reference function.~\cite{frozen-pccd-jcp-2014,pccd-tailored-jctc-2022}
In general, a tCC wave function has the ansatz
\begin{equation}\label{eq:tcc}
\ket{{\rm tCC}} = e^{\hat{T}} \ket{\Phi_0} = e^{\hat{T}^{\rm ext}} e^{\hat{T}^{\rm int}} \ket{\Phi_0},
\end{equation}
where $\ket{\Phi_0}$ is a reference Slater determinant and $\hat{T}$ is the cluster operator that is partitioned into a sum of two disjoint cluster operators, $\hat{T}^{\rm int}$ and $\hat{T}^{\rm ext}$.
By construction, $\hat{T}^{\rm int}$ and $\hat{T}^{\rm ext}$ commute if the cluster operators are pure particle-hole excitation operators and the above equation is valid.
In pCCD-tCC,~\cite{frozen-pccd-jcp-2014,pccd-tailored-jctc-2022,pccp-geminal-review-pccp-2022} the cluster operator $\hat{T}^{\rm int}$ in eq.~\eqref{eq:tcc} is taken as the pCCD pair-excitation-only cluster operator (see eq.~\eqref{eq:pccd}).
Since pCCD describes only electron-pair excitations, the cluster amplitudes in the tCC ansatz are thus divided into the conventional electron-pair amplitudes ($\hat{T}^{\rm int}$) and non-pair (or broken-pair) amplitudes ($\hat{T}^{\rm ext}$).
The pCCD-tCC ansatz, therefore, reads
\begin{equation}
\ket{\mathrm{pCCD-tCC}} = e^{\hat{T}^{\rm ext}} \ket{{\rm pCCD}} = e^{\hat{T}^{\rm ext}} e^{\Tp} \ket{\Phi_0}.
\end{equation}
From an implementation perspective, the pCCD-tCC model is easily obtained from a conventional CC implementation by \textit{freezing} the pair amplitudes.
The \textit{freezing} step requires the pCCD electron-pair amplitudes as input data and neglects those pair contributions to the CC vector function (setting them to zero).
Thus, pCCD-tCC is commonly known as the frozen-pair (fp)CC ansatz.~\cite{frozen-pccd-jcp-2014}
An alternative CC correction on top of pCCD exploits a linearized (L)CC ansatz,~\cite{ap1rog-lcc-jctc-2015, diagrammtic-lcc-carter-fenk-jpca-2025} where the corresponding wavefunction is approximated as
\begin{equation}
    |{\text{fpLCC}}\rangle \approx \left(1 + \hat T^{\rm ext} \right) |{\text{pCCD}}\rangle.
    \label{eq:fpLCCD}
\end{equation}
Technically, the above ansatz does not correspond to a linearized tCC flavor as the electron-pair contributions (from the pCCD reference state) are not linearized.
Instead, fpLCC can be understood as a linearized pCCD-tCC approach, where the full frozen-pair equations (including all non-linear and disconnected terms involving $\Tp$ vertices) are included in the LCC equations.

Most frequently, the $\hat{T}^{\rm ext}$ cluster operator contains at most (broken-pair) double excitations.
For instance, the cluster operator of fpCCD is defined as $\hat{T}^{\rm ext} = \hat{T}_2^\prime = \hat T_2 - \Tp$, while the cluster operator of fpCCSD includes also single excitations, $\hat{T}^{\rm ext} = \hat{T}_1 + \hat{T}_2^\prime$.
In the following, we restrict all CC equations to the spin-free case, where all excitation operators are written in terms of singlet excitation operators $\hat E_a^i$,
\begin{equation}
\label{eq:singlet_excitation_op}
    \hat E_a^i = \hat a_{a}^\dagger \hat a_{i} + \hat a_{\bar a}^\dagger \hat a_{\bar i}.
\end{equation}
For spin-free double excitations, the $\hat T_2$ cluster operator takes on the form
\begin{equation}
    \hat T_2 = \frac{1}{2} \sum_{i j a b} t_{ij}^{ab} \, \hat E_{a}^{i} \hat E_{b}^{j},
    \label{eq:ccd_spinadapted}
\end{equation}
while single excitations reduce to
\begin{equation}
    \hat T_1 = \sum_{i a} t_{i}^{a} \, \hat E_{a}^{i}.
    \label{eq:ccs_spinadapted}
\end{equation}
The correlation energy of the fpCC model can be obtained by solving the corresponding time-independent projected Schr\"odinger equation
\begin{align}\label{eq:fpcc-se}
    \bra{K} e^{-\Tp} e^{-\hat T^{\rm ext}} \hat{H}_N e^{\hat{T}^{\rm ext}} e^{\Tp} \ket{\Phi_0} &= 0,
\end{align}
where $\bra{K}$ are all singly, doubly-, etc.~excited determinants that can be generated by the external excitation operator $\hat{T}^{\rm ext}$ acting on the pCCD reference state $\ket{\Phi_0}$ and $\hat{H}_N $ is the molecular Hamiltonian in its normal-product form.
Specifically, for the fpCC model, we can define the fpCC-similarity transformed Hamiltonian in normal-product form as
\begin{align}\label{eq:fpcc-effham}
    \hat{\mathcal{H}}_N^{\mathrm{(fpCC)}}
    &= e^{-\Tp} e^{-\hat T^{\rm ext}} \hat{H}_N e^{\hat{T}^{\rm ext}} e^{\Tp} \nonumber \\
    &= {\hat{H}_N}  + [\hat{H}_N,\Tp] + \frac{1}{2} [[\hat{H}_N,\Tp],\Tp] + \ldots \nonumber \\
    &\phantom{=+} + [\hat{H}_N,\Te] + \frac{1}{2} [[\hat{H}_N,\Te],\Te] + \ldots \nonumber \\
    &\phantom{=+} +                               [[\hat{H}_N,\Te],\Tp] + \frac{1}{2} [[[\hat{H}_N,\Te],\Tp],\Tp] + \ldots
\end{align}
For the linearized version, the fpLCC-similarity transformed Hamiltonian in normal-product form neglects all non-linear terms involving pure $\Te$ terms, while all $\Tp$ terms as well as the coupling term between $\Tp$ and $\Te$ survive (indicated by the subscript $L$ below),  
\begin{align}\label{eq:fplcc-effham}
    \hat{\mathcal{H}}_N^{\mathrm{(fpLCC)}}
    &= \left [ e^{-\Tp} e^{-\hat T^{\rm ext}} \hat{H}_N e^{\hat{T}^{\rm ext}} e^{\Tp} \right ]_L\nonumber \\
    &= {\hat{H}_N} + [\hat{H}_N,\Tp] + \frac{1}{2} [[\hat{H}_N,\Tp],\Tp] + \ldots \nonumber \\
    &\phantom{=+} + [\hat{H}_N,\Te] \nonumber \\
    &\phantom{=+} + [[\hat{H}_N,\Te],\Tp] + \frac{1}{2} [[[\hat{H}_N,\Te],\Tp],\Tp] + \ldots
\end{align}

Numerical studies suggest~\cite{ap1rog-lcc-jctc-2015,pccd-ptx-jctc-2017} that fpCC and fpLCC are reliable wave function ans\"atze to model both static and dynamic electron correlation, approaching chemical accuracy ($\sim1$ kcal/mol) for many challenging systems.~\cite{pccd-ptx-jctc-2017}
Originally formulated for ground-state electronic structure problems, pCCD-tCC methods need to be extended to target electronically excited states.
Possible extensions are the equation-of-motion (EOM) formalism,~\cite{ eom-cc-rowe-rmp-1968, eom-cc-cpl-1989, eom-ccsd-bartlett-jcp-1993} linear-response theory,~\cite{lr-eom-ccc-sekino-ijqc-1984, koch-cc-response-jcp-1990, koch-lr-ccsd-jcp-1990, christiansen-response-review-ijqc-1998} or state-specific calculations.
Within the EOM formalism, excited-state electronic structures have been treated either at the pCCD level, where single excitations are accounted for approximately,~\cite{eom-pccd-jcp-2016, eom-pccd-erratum-jcp-2017} or for an fpCC reference function,~\cite{eom-pccd-lccsd-jctc-2019,pccd-tailored-bartlett-jcp-2023} which also accounts for dynamical correlation effects.
In this work, we will focus on various EOM models for an fpCC reference state.
\subsection*{Extending frozen-pair CC methods to model excited states}\label{sec:theory}
Within the EOM formalism,~\cite{ eom-cc-rowe-rmp-1968, eom-cc-cpl-1989, eom-ccsd-bartlett-jcp-1993} excited states are modelled using a linear CI-type ansatz,
\begin{equation}\label{eq:eom-ci}
        \hat{R} = \sum_\mu c_\mu \hat{\tau}_\mu,
\end{equation}
where the sum runs over all (spin-free) excitations present in the (spin-free) cluster operator as well as the identity operator $\hat{\tau}_0$, accounting for ground-state contributions to the targeted excited state.
The excited state is obtained by $\hat{R}$ acting on the CC reference state,
\begin{equation}
        \ket{\Psi} = \hat{R} e^{\Tt} \ket{\Phi_0} = \sum_\mu {c_\mu \hat{\tau}_\mu} e^{\Tt} \ket{\Phi_0}.
\end{equation}
Using the wavefunction ansatz above, we can derive the well-known EOM-CC equations,
\begin{equation}\label{eq:eomcc}
        [\hat{\mathcal{H}}_N^{(\rm CC)},\hat{R}] \ket{\Phi_0}  = \omega \hat{R} \ket{\Phi_0},
\end{equation}
containing the similarity-transformed Hamiltonian in normal-product form for a given CC model, and the excitation energies $ \omega=(E-E_0)$ with respect to the CC ground state.

Since we will focus on excitation energies, we have to solve for the $\hat{R}$ amplitudes only.
The corresponding excited states and excitation energies are obtained by diagonalizing the $\hat{\mathcal{H}}_N^{(\rm CC)}$ represented in the configurational space generated by the singly-, doubly-, etc. excitation operators $\hat \tau_\mu$ included in eq.~\eqref{eq:eom-ci}.
For a conventional CC reference state (like CCSD, CCSDT, etc.), the non-Hermitian Hamiltonian matrix to be diagonalized has the form
\begin{equation}
\bm H^{\rm EOM-CCSD\ldots} =
        \begin{bmatrix}                                                         
          0 & \bra{0} \Hn \ket{S} & \bra{0} \Hn \ket{D} & \ldots \\
          0 & \bra{S} \Hn \ket{S} & \bra{S} \Hn \ket{D} & \ldots\\
          0 & \bra{D} \Hn \ket{S} & \bra{D} \Hn \ket{D} & \ldots\\            
     \vdots &      \vdots         &      \vdots         & \ddots\\            
        \end{bmatrix},                                                     
\end{equation}                                                                  
where $\ket{S}$ indicates the single-excitation manifold (generated by $\Tt_1\ket{\Phi_0}$), $\ket{D}$ labels the double-excitation manifold (generated by $\Tt_2\ket{\Phi_0}$), etc.
If $\Hn$ of the above equation is replaced by the pCCD-tCC counterpart, the corresponding diagonalization problem has to be modified,~\cite{pccd-tailored-bartlett-jcp-2023}
\begin{equation}\label{eq:hptcc}
\bm H^{\rm EOM-ptCCSD\ldots} =
        \begin{bmatrix}                                                         
          0         & \bra{0} \Hfp \ket{S} & \bra{0} \Hfp \ket{P} & \bra{0} \Hfp \ket{D} & \ldots \\
          0         & \bra{S} \Hfp \ket{S} & \bra{S} \Hfp \ket{P} & \bra{S} \Hfp \ket{D} & \ldots\\
\bra{P}\Hfp\ket{0}  & \bra{P} \Hfp \ket{S} & \bra{P} \Hfp \ket{P} & \bra{P} \Hfp \ket{D} & \ldots\\
          0         & \bra{D} \Hfp \ket{S} & \bra{D} \Hfp \ket{P} & \bra{D} \Hfp \ket{D} & \ldots\\            
     \vdots         &      \vdots          &      \vdots          &      \vdots          & \ddots\\            
        \end{bmatrix},                                                     
\end{equation}
where a non-vanishing term in the first column arises as the electron-pair excitation amplitudes do not satisfy the projected Schr\"odinger equation for the pCCD-tCC problem (the $\Tp$ amplitudes satisfy only $\bra{P}\Hp\ket{0} = 0$, while $\bra{P}\Hfp\ket{0} \neq 0$).
This issue can be resolved if $\Hn$ in the diagonalization problem is adjusted so that pair and broken-pair amplitudes are treated differently.
If we enforce the projection equations to be satisfied for each excitation manifold, the Hamiltonian matrix to be diagonalized becomes 
\begin{equation}\label{eq:hfpcc}
\bm H^{\rm EOM-fpCCSD\ldots} =
        \begin{bmatrix}                                                         
          0         & \bra{0} \Hfp \ket{S} & \bra{0} \Hfp \ket{P} & \bra{0} \Hfp \ket{D} & \ldots \\
          0         & \bra{S} \Hfp \ket{S} & \bra{S} \Hfp \ket{P} & \bra{S} \Hfp \ket{D} & \ldots\\
          0         & \bra{P} \Hp  \ket{S} & \bra{P} \Hp  \ket{P} & \bra{P} \Hp  \ket{D} & \ldots\\
          0         & \bra{D} \Hfp \ket{S} & \bra{D} \Hfp \ket{P} & \bra{D} \Hfp \ket{D} & \ldots\\            
     \vdots         &      \vdots          &      \vdots          &      \vdots          & \ddots\\            
        \end{bmatrix}.                                            
\end{equation}
Thus, the first column ``vanishes'' as the pCCD amplitude equations ($\bra{P}\Hp\ket{0} = 0$) are satisfied, while all broken-pair amplitudes are optimized within the fpCC step.
For the linearized fpCC variant,~\cite{eom-pccd-lccsd-jctc-2019} we have to replace $\Hfp$ in eq.~\eqref{eq:hfpcc} with $\Hfpl$ defined in eq.~\eqref{eq:fplcc-effham}.
In all investigated EOM-fpCC models, we restrict the CC reference state to at most double excitations, yielding the corresponding EOM-ptCCSD (see eq.~\eqref{eq:hptcc}) or EOM-fpCCSD (see eq.~\eqref{eq:hfpcc}) models.
All working equations are summarized in the SI.

Finally, some of us also developed a very simple EOM extension for a pCCD reference state, which was inspired by configuration interaction singles (CIS).~\cite{eom-pccd-jcp-2016,eom-pccd-erratum-jcp-2017,pccd-ee-f0-actinides}
In the corresponding excited-state model, both electron-pair (as in the ground-state reference function) and single excitations are included,
\begin{equation}\label{eq:eom-pccd+s-r-operator}
        \hat{R} = \sum_{\mu \in 0,S,P} c_\mu \hat{\tau}_\mu,
\end{equation}
while the CC ground-state reference function is restricted to the pCCD state.\cite{eom-pccd-jcp-2016,eom-pccd-erratum-jcp-2017}
The corresponding diagonalization problem has a rather simple form and includes only single and electron-pair excited determinants in the projection manifold,
\begin{equation}\label{eq:hpccd}
\bm H^{\rm EOM-pCCD+S\ldots} =
        \begin{bmatrix}                                                         
          0         & \bra{0} \Hp \ket{S} & \bra{0} \Hp \ket{P} \\
\bra{S} \Hp \ket{0} & \bra{S} \Hp \ket{S} & \bra{S} \Hp \ket{P} \\
          0         & \bra{P} \Hp \ket{S} & \bra{P} \Hp  \ket{P} \\
        \end{bmatrix}.                                            
\end{equation} 
We should stress again that we work in a spin-free picture, where only singlet excitations are accessible within all EOM formulations.
The corresponding projection manifolds contain the spin-free singly and spin-free doubly excited determinants with respect to the CC reference state $\ket{\Phi_0}$ (see also Refs.~\citenum{helgaker-book-2000,ap1rog-lcc-jctc-2015}).
Finally, both the EOM-ptCCSD and EOM-fpCCSD models have the same computational scaling as the EOM-CCSD counterpart, which is limited by the $\mathcal{O}(N^6)$ bottleneck.

\begin{figure}[t] 
    \centering
    \includegraphics[width=1.0\textwidth]{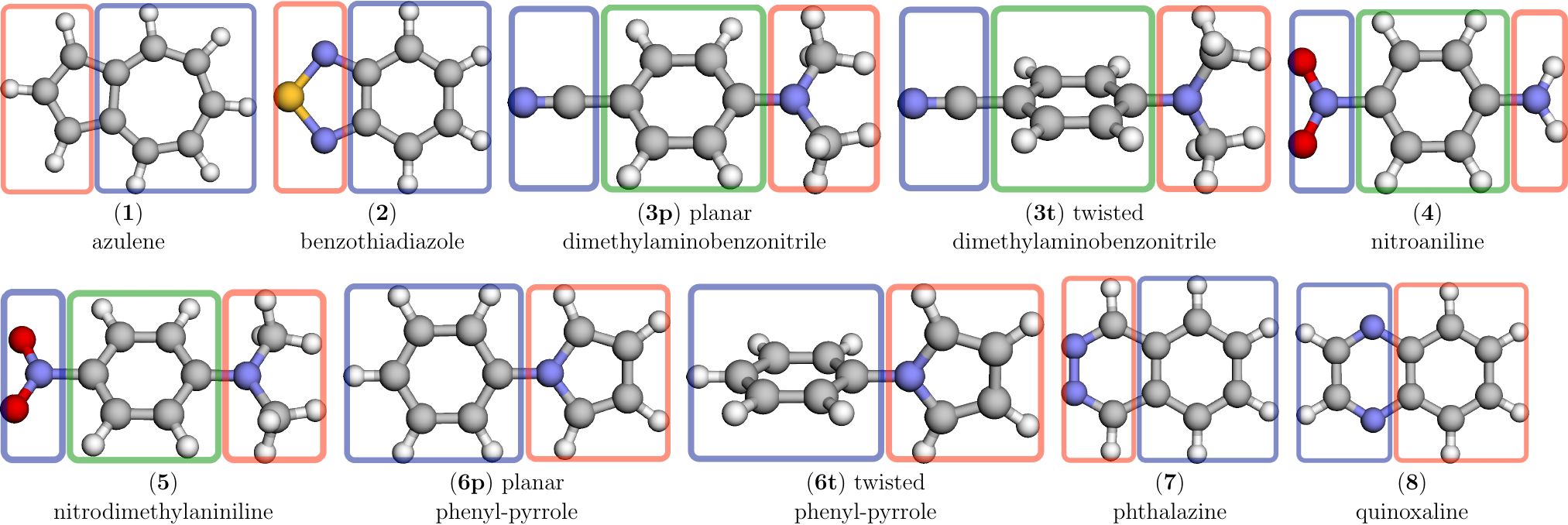} 
    \caption{Molecular data set with significant CT states.
    Red: donor domain.
    Green: bridge.
    Blue: acceptor domain.
    The molecular coordinates have been taken from the QUEST database.~\cite{quest-jctc-2025}
    All molecules are visualized using the PyBEST GUI.~\cite{pybest-gui-ijqc-2025}
    }\label{fig:structures-ct}
\end{figure}
\begin{figure}[t] 
    \centering
    \includegraphics[width=1.0\textwidth]{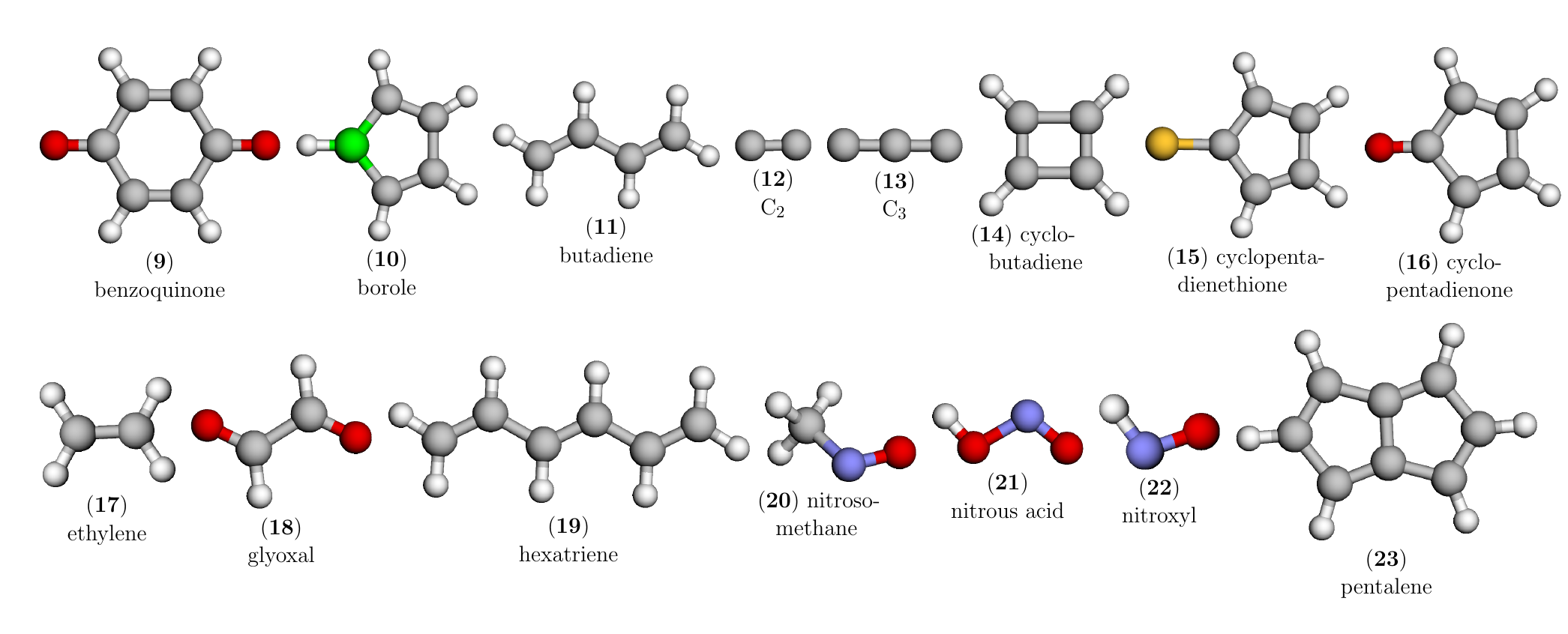} 
    \caption{Molecular data set with dominant doubly-excited state character.
    The molecular coordinates have been taken from the QUEST database.~\cite{quest-jctc-2025}
    All molecules are visualized using the PyBEST GUI.~\cite{pybest-gui-ijqc-2025}
    }\label{fig:structures-double-excitations}
\end{figure}
\section*{Computational Details}
All EOM-CCSD, EOM-fpCCSD, and EOM-ptCCSD calculations were performed in a developer version of the PyBEST-2.2.0.dev0 software package~\cite{pybest-paper-cpc-2021, pybest-paper-update1-cpc-2024} using the cc-pVDZ, aug-cc-pVDZ, cc-pVTZ, and aug-cc-pVTZ basis sets.~\cite{cc-pvxz-dunning-jcp-1989, aug-cc-pvxz-jcp-1992}
The calculations for doubly excited states used the aug-cc-pVTZ basis set solely to match the best theoretical estimates from the literature.

We used the Cholesky decomposed two-electron integrals~\cite{cholesky-review-2011} with a threshold set to 10$^{-4}$, which is sufficient for excitation/relative energies.~\cite{ea-eom-pccd-frozen-pair-jctc-2025}
Core orbitals were kept frozen in all correlated calculations.
We GPU-accelerated our CC and EOM-CC calculations on NVIDIA H100 cards according to PyBEST implementations.~\cite{pybest-gpu-jctc-2024, pybest-gpu-h100-gh200-cupy-pytorch-jctc-2026}
EOM calculation used canonical Hartree--Fock orbitals and variationally optimized natural pCCD orbitals~\cite{oo-ap1rog-prb-2014, ap1rog-non-variational-orbital-optimizarion-jctc-2014, tamar-pccd-jcp-2014} and are denoted in parentheses as HF and pCCD, respectively, such as EOM-CCSD(HF) vs.~EOM-CCSD(pCCD).
In our excited-state calculations, we used the Davidson diagonalization and computed up to 15 lowest-lying singlet-singlet states for each EOM-CC method. 

To access the charge-transfer character of the electronic excited states, we used the recently introduced fully automated domain-based charge-transfer implementations~\cite{daispy-domain-ct-benchmark-2026} in PyBEST as present in the DAISpY tool~\cite{daispy-domain-ct-code-2026} (see Refs.~\cite {eom-pccd+s-ct-analysis-jctc-2025, eom-pccd+s-ct-analysis-carbazole-ram-jpca-2025} for examples). 
In a nutshell, the molecule is divided into chemically relevant regions (domains), and excitations are analyzed in terms of electron transfer between these domains (see also Figure~\ref{fig:structures-ct} for a graphical representation of the domains).
Such a framework is applicable for any CI-type excited-state wavefunction, including EOM-CC methods.
Furthermore, we analyze the directed CT character (dCT), which, for a donor--bridge--acceptor pathway as visualized in Figure~\ref{fig:structures-ct}, is obtained from the individual domain-based pathways ($(D \to B)$ or $(B \to A)$) as
\begin{align}\label{eq:dct}
\mathrm{dCT}(D \to B \to A) =\, 
    &\left[(D \to B) + (B \to A) + (D \to A)\right] - \left[(D \leftarrow B) + (B \leftarrow A) + (D \leftarrow A)\right].
\end{align}
For a two-domain pathway, we only consider the donor--acceptor pathways, namely $(D \to A)$ and $(D \leftarrow A)$.
The DAISpY analysis is performed for the weighted domain-accumulation strategy only.~\cite{daispy-domain-ct-code-2026}

\section*{Results and Discussion}
In the following, we will scrutinize the performance of our newly introduced EOM-fpCCSD model by comparing it against the EOM-ptCCSD and EOM-CCSD models and reference theoretical data for challenging excited states.
We divide our discussion into two subsections: CT states and doubly excited states. 
\subsection*{CT Excitations}
We start our evaluation of the EOM-fpCCSD ansatz by calculating the lowest-lying singlet–singlet vertical excitation energies of the molecules depicted in Figure~\ref{fig:structures-ct}.
These molecular systems form a charge-transfer subset of the QUEST database.~\cite{quest-jctc-2025}

Table~\ref{tab:ee-eom} summarizes the low-lying vertical excitation energies obtained using various EOM-CC models, along with their CT character (see the Computational Details section for further information).
Entries marked ‘nc’ indicate states that could not be computed within a given method.
Each excited state is labeled according to the QUEST database notation, where ‘Val’, ‘CT’, and ‘Ryd’ denote valence, charge-transfer, and Rydberg states, respectively.
The symbol ‘dCT’ in the third column indicates the strength of the directed charge-transfer character: 0 corresponds to no CT, while $+$, $++$, and $+++$ denote increasing degrees of CT strength. 
The percentage contribution of this directed CT is given by the ‘dCT[\%]’ parameter (defined in eq.~\eqref{eq:dct}), with positive values corresponding to net donor-to-acceptor ($D \rightarrow A$) transitions and negative values to net acceptor-to-donor ($D \leftarrow A$) transitions (cf. Computational Details section for more details). 
The donor, bridge, and acceptor domains are indicated in Figure~\ref{fig:structures-ct} by red, green, and blue fragments, respectively.

\begin{figure}[t] 
    \centering
    \includegraphics[width=0.5\textwidth]{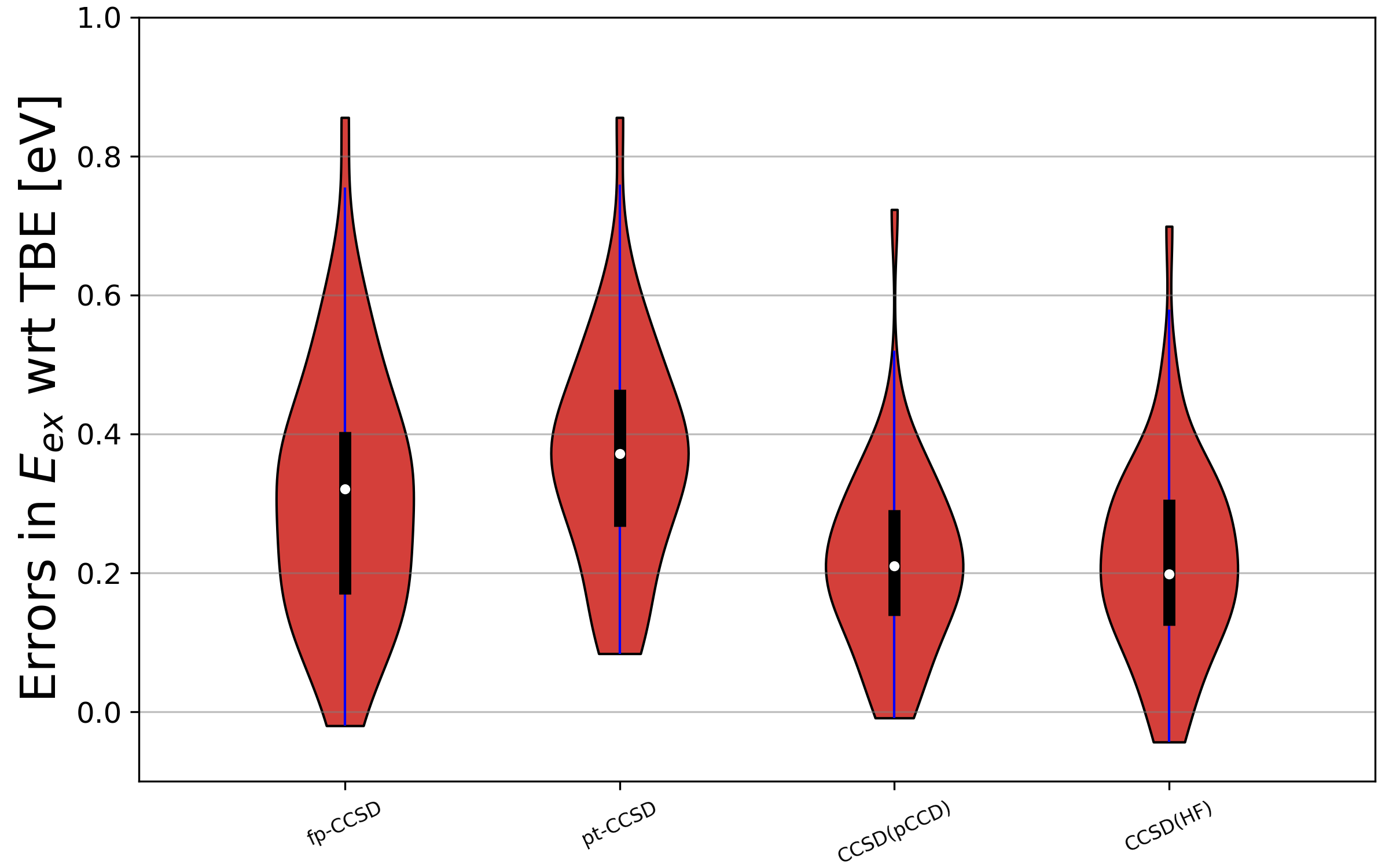} 
    \caption{Violin plots comparing the performance of different EOM-CC models for excitation energies against the theoretical best estimate (TBE). The shape of each violin illustrates the distribution of errors: its width reflects data density (wider = higher density), and its height spans the full data range. The white dot indicates the median, and the thick bar denotes the interquartile range (IQR).
    CCSD(pCCD): EOM-CCSD exploiting pCCD-optimized natural orbitals.
    CCSD(HF): EOM-CCSD using canonical Hartree--Fock orbitals.
    }\label{fig:violin}
\end{figure}

Since the CT analysis within the EOM-CC methods is more convenient in the (localized) pCCD natural orbital basis, we first compare the excitation energies from standard EOM-CCSD using pCCD natural orbitals versus canonical Hartree–Fock orbitals.
As shown in Table~\ref{tab:ee-eom}, both orbital bases yield very similar excitation energies for the EOM-CCSD method, with deviations typically within 0.01 eV. 
They also exhibit very similar statistical errors with respect to the theoretical best estimates, with a ME/MAE of approximately 0.22 eV and a SD of 0.13-0.14 eV.
The EOM-fpCCSD spectra closely resemble those of EOM-CCSD; however, the statistical errors relative to the TBE are slightly larger.
The largest MEs and MAEs from TBE are observed for the EOM-ptCCSD model. 
However, its SD is smaller then EOM-fpCCSD. 
These statistical differences are best illustrated using violin plots presented in Figure~\ref{fig:violin}. 

\begin{figure}[t] 
    \centering
    \includegraphics[width=0.5\textwidth]{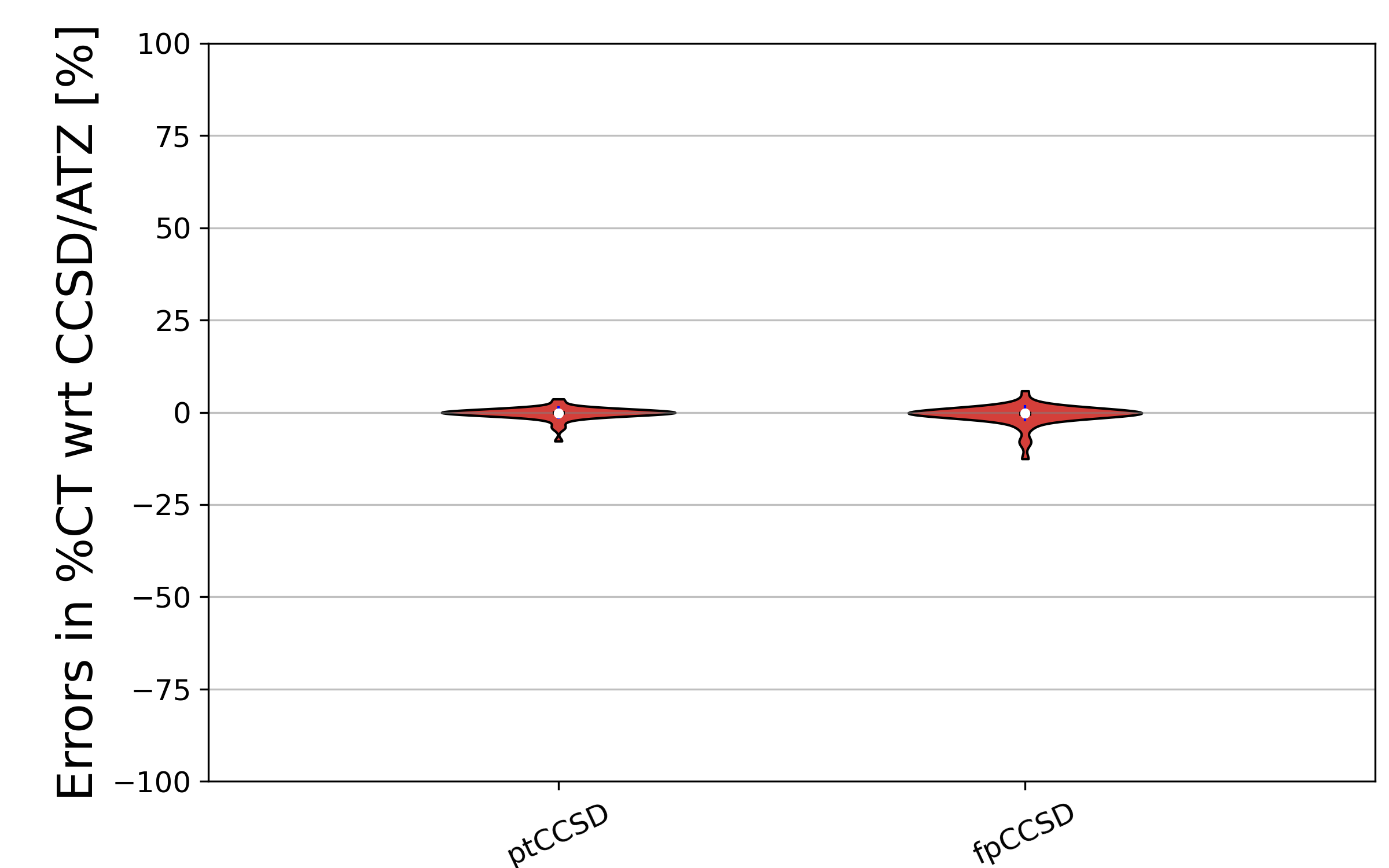} 
    \caption{Violin plot for the method dependence of the directed CT character with respect to EOM-CCSD(pCCD) for the aug-cc-pVTZ (ATZ) basis set.
    All EOM-CC calculations are done with pCCD-optimized natural orbitals.
    The shape of each violin illustrates the distribution of errors: its width reflects data density (wider = higher density), and its height spans the full data range. The white dot indicates the median, and the thick bar denotes the interquartile range (IQR).
    }\label{fig:ct_error_wrt_ccsd}
\end{figure}

The CT analysis yields highly consistent results across the EOM-CCSD, EOM-ptCCSD, and EOM-fpCCSD methods for all investigated systems and excited states.
As shown in Figure~\ref{fig:ct_error_wrt_ccsd}, the dCT is largely method-independent.
A detailed comparison in Table~\ref{tab:ee-eom} reveals that the sign of the directed CT is identical for nearly all states, while the dCT[\%] values typically agree within 1–2\%.
The only notable exception is structure (\textbf{6t}), where larger discrepancies are observed.
The weakest directed CT character is found for structure (\textbf{1}), whereas the strongest CT character appears in structures (\textbf{3t}), (\textbf{4}), (\textbf{5}), (\textbf{6}), (\textbf{7}), and (\textbf{8}).
These results are in good agreement with the CT classification provided in the QUEST database.
Interestingly, our analysis also identifies a non-negligible CT admixture in several Rydberg and valence states that are formally classified as non-CT in the QUEST database.
The corresponding directed CT character ranges from 10\% (negligible) to 35\% (considerable) or even 60\% (lone-pair to $\pi^*$ excitations).
This highlights a key advantage of our approach: it enables a quantitative and method-consistent assessment of the strength of individual CT contributions.
The corresponding numerical values are provided in the ESI.

\begin{figure}[t] 
    \centering
    \includegraphics[width=1.0\textwidth]{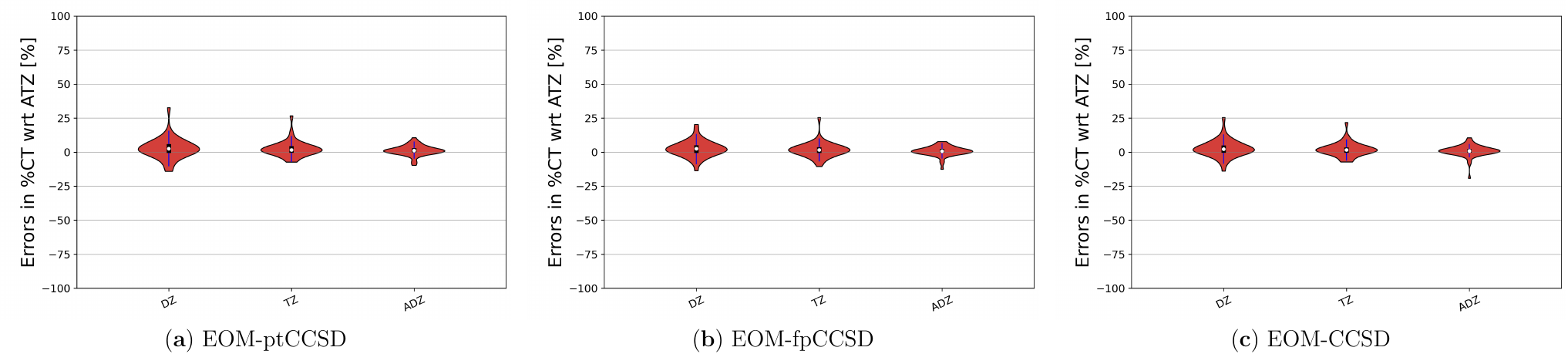} 
    \caption{Violin plot for the basis set dependence of the directed CT character.
    All errors within a method are determined with respect to the aug-cc-pVTZ basis set.
    All EOM-CC calculations are done with pCCD-optimized natural orbitals.
    The shape of each violin illustrates the distribution of errors: its width reflects data density (wider = higher density), and its height spans the full data range. The white dot indicates the median, and the thick bar denotes the interquartile range (IQR).
    }\label{fig:ct_basis_set_error}
\end{figure}

All investigated EOM-CC methods show a comparable basis-set dependence for the computed excitation energies, with detailed numerical data provided in the ESI.
The largest basis-set effects are observed for Rydberg-type excited states, especially when augmented functions are included.
We should stress that the dCT character features a strong basis-set resistance (see Figure~\ref{fig:ct_basis_set_error}), with an average SD of 5\% across all investigated methods.
As shown in the figure, the use of pCCD natural orbitals significantly reduces the sensitivity of the dCT to the basis-set size. 
A similar observation has been made for pCCD-based ionization potentials and electron affinities.~\cite{ip-eom-pccd-chem-comm-2021, ea-eom-pccd-jpca-2024, pccd-ip-ea-mod-koopmans-jcp-2025, ea-eom-pccd-frozen-pair-jctc-2025, dip-fpccsd-jctc-2025, pccd-ekt-jctc-2026}
Larger errors (for cc-pVDZ and cc-pVTZ compared to their augmented counterparts) are only observed if Rydberg-type contributions have to be considered in the excited states of interest.

\begin{table}[tb]
    \centering
    \resizebox{1\textwidth}{!}{
    \begin{tabular}{c l c c c c c c c c c}
    \hline
    Mol& State & dCT
                         & EOM-CCSD(HF) &  \multicolumn{2}{c}{EOM-CCSD(pCCD)} 
                         &  \multicolumn{2}{c}{EOM-ptCCSD (pCCD)} 
                         &  \multicolumn{2}{c}{EOM-fpCCSD (pCCD)}
                         & TBE~\cite{quest-jctc-2025}
                         \\
    &   &   type         & $E_{\rm ex}$ [eV] & $E_{\rm ex}$ [eV] &  {dCT[\%]}
                         & $E_{\rm ex}$ [eV] &  {dCT[\%]}
                         & $E_{\rm ex}$ [eV] &  {dCT[\%]}
                         & $E_{\rm ex}$ [eV]
                         \\
    \hline
    \multirow{6}{*}{(\textbf{1})}&
B2 (Val, $\pi$-$\pi^*$)   & +  & 2.278 & 2.333 & 14.8 & 2.600 & 14.2 & 2.374 & 14.7 & 2.162\\
&A1 (CT, $\pi$-$\pi^*$)   & +  & 3.973 & 3.984 & 18.8 & 4.186 & 18.9 & 4.170 & 19.1 & 3.849\\
&B2 (CT, $\pi$-$\pi^*$)   & 0  & 4.780 & 4.790 &  9.7 & 4.983 &  8.0 & 4.857 &  9.5 & 4.510\\
&A2 (Ryd, n.d.)           & 0  & 4.903 & 4.913 &  9.6 & 4.990 &  8.8 & 4.990 &  8.8 & 4.874\\
&A1 (Val, $\pi$-$\pi^*$)  & +  & 5.216 & 5.213 & 22.9 & 5.363 & 23.4 &  nc   &  nc  & 4.956\\
&B1 (Ryd, n.d.)           & 0  & 5.312 & 5.326 &  8.2 & 5.407 &  7.6 & 5.407 &  7.6 & 5.302
\\ \hline
    \multirow{4}{*}{(\textbf{2})}&
B2 (CT, $\pi$-$\pi^*$)      & +  &  4.559 & 4.556 & 21.4  & 4.771 & 21.9  & 4.531 & 22.2  & 4.291\\
&A1 (Val, $\pi$-$\pi^*$)    & +  &  4.540 & 4.600 & -19.8 & 4.837 & -19.9 & 4.817 & -20.2 & 4.371\\
&A2 (Val, $n$-$\pi^*$)      & ++ &  5.019 & 5.073 & 34.4  & 5.232 & 34.2  & 5.232 & 34.2  & 4.806\\
&B1 (Val, $n$-$\pi^*$)      & +  &  5.694 & 5.727 & 11.7  & 5.876 & 11.6  & 5.876 & 11.6  & 5.422
\\ \hline
    \multirow{4}{*}{(\textbf{3p})}&
B2 (Val, $\pi$-$\pi^*$)         & +  &  4.534 & 4.535 &  12.1 & 4.630 &  12.5 & 4.483 &  11.1 & 4.336\\
&B1 (Ryd, n.d.)                 & +  &  4.915 & 4.914 & -13.0 & 4.938 & -12.7 & 4.938 & -12.7 & 4.816\\
&A1 (Val, CT, $\pi$-$\pi^*$)    & ++ &  5.023 & 5.024 &  29.4 & 5.122 &  30.0 & 5.080 &  29.5 & 4.866\\
&A2 (Ryd, n.d.)                 & 0  &  5.551 & 5.551 &  -6.5 & 5.577 &  -6.1 & 5.577 &  -6.1 & 5.457
\\ \hline
    \multirow{2}{*}{(\textbf{3t})}&
A2 (Val, CT, $n$-$\pi^*$)   &+++& 4.354 & 4.361 & 70.3 & 4.465 & 70.0 & 4.465 & 70.0 & 4.117\\
&B1 (Val, CT, $n$-$\pi^*$)  &+++& 5.092 & 5.092 & 69.5 & 5.187 & 69.3 & 5.187 & 69.3 & 4.755
\\ \hline
    \multirow{4}{*}{(\textbf{4})}&
A2 (Val, $n$-$\pi^*$)            & 0 & 4.160 & 4.170 & -5.5 & 4.361 & -5.6 & 4.361 & -5.6 & 3.998\\
&A1 (Val, CT, $\pi$-$\pi^*$)     &+++& 4.634 & 4.635 & 50.9 &  nc   & nc   & 4.732 & 43.3 & 4.402\\
&B2 (Val, $\pi$-$\pi^*$)         & + & 4.692 & 4.691 & 11.8 &  nc   & nc   & 4.598 & 17.6 & 4.508\\
&B1 (Val, $n$-$\pi^*$)           & 0 & 4.694 & 4.705 & -3.7 & 4.899 & -3.9 & 4.899 & -3.9 & 4.507

\\ \hline
    \multirow{1}{*}{(\textbf{5})}&
A1 (Val, CT, $\pi$-$\pi^*$)  &+++&4.386 & 4.385 & 55.0 & 4.533 & 53.5 & 4.458 & 51.1 & 4.133
\\ \hline
    \multirow{5}{*}{(\textbf{6p})}&
 B2 (Val, $\pi$-$\pi^*$)     & 0 & 4.869 & 4.867 &  6.5 & 4.984 &  7.5 & 4.736 &  5.6 & 4.695\\
&A1 (Val, $\pi$-$\pi^*$)     & 0 & 5.150 & 5.149 &  2.6 & 5.274 &  2.8 & 5.178 &  2.5 & 5.003\\
&B2 (Val, CT, $\pi$-$\pi^*$) & ++& 5.628 & 5.652 & 35.6 & 5.844 & 34.9 & 5.835 & 34.6 & 5.346\\
&B1 (Ryd, $\pi$-$3s$)        & 0 & 5.647 & 5.633 &  9.8 & 5.658 & 10.1 & 5.658 & 10.1 & 5.574\\
&A2 (Ryd, $\pi$-$3s$)        & ++& 5.751 & 5.764 & 31.9 & 5.874 & 31.1 & 5.874 & 31.1 & 5.628
\\ \hline
    \multirow{6}{*}{(\textbf{6t})}&
 B2 (Val, $\pi$-$\pi^*$)     & 0 &5.197 & 5.198 &  1.6 & 5.335 &  1.4 & 5.021 &  1.0 & 5.041\\
&A2 (Ryd, n.d.)              & + &5.627 & 5.638 & 24.1 & 5.747 & 23.8 & 5.747 & 23.8 & 5.524\\
&B2 (Val, CT, $\pi$-$\pi^*$) &+++&5.954 & 5.977 & 63.6 & 6.267 & 55.9 & 6.260 & 55.1 & 5.615\\
&A1 (Val, CT, $\pi$-$\pi^*$) &+++&5.999 & 6.031 & 51.0 & 6.196 & 46.6 & 6.086 & 38.4 & 5.657\\
&B1 (Ryd, n.d.)              & ++&6.078 & 6.095 & 30.8 & 6.228 & 32.3 & 6.223 & 31.9 & 5.946\\
&A2 (Val, CT, $\pi$-$\pi^*$) &+++&6.258 & 6.242 & 68.1 & 6.325 & 67.8 & 6.324 & 67.8 & 5.949
\\ \hline
    \multirow{11}{*}{(\textbf{7})}&
A2 (Val, wCT, $n$-$\pi^*$)   & + & 4.247 & 4.261 &  17.7 & 4.433 & 17.0 & 4.433 & 17.0 & 3.898\\
&B1 (Val, wCT, $n$-$\pi^*$)  & ++& 4.608 & 4.615 &  32.6 & 4.763 & 32.8 & 4.763 & 32.8 & 4.302\\
&A1 (Val, $\pi$-$\pi^*$)     & 0 & 4.642 & 4.672 &  -4.3 & 4.844 & -4.5 & 4.704 & -4.5 & 4.451\\
&B2 (Val, $\pi$-$\pi^*$)     & + & 5.368 & 5.406 & -11.8 & 5.678 &-12.7 & 5.657 &-13.0 & 5.188\\
&B1 (Val, wCT, $n$-$\pi^*$)  & + & 5.908 & 5.930 &  17.0 & 6.120 & 16.5 & 6.120 & 16.5 & 5.548\\
&A2 (Val, $n$-$\pi^*$)       &+++& 6.541 & 6.565 &  63.0 & 6.698 & 63.8 & 6.698 & 63.8 & 5.842\\
&A2 (Val, wCT, $n$-$\pi^*$)  & ++& 6.224 & 6.235 &  31.5 & 6.393 & 31.8 & 6.393 & 31.8 & 5.876\\
&A1 (Val, $\pi$-$\pi^*$)     & 0 & 6.422 & 6.409 &  -4.0 & 6.599 & -7.8 & 6.324 & -7.8 & 6.109\\
&B2 (Ryd, n.d.)              & ++& 6.546 & 6.550 &  34.7 & 6.610 & 37.4 & 6.608 & 36.8 & 6.263\\
&A1 (Val, $\pi$-$\pi^*$)     & 0 & 6.650 & 6.636 &  -6.3 & 6.794 & -2.8 & 6.765 & -3.4 & 6.408\\
&A2 (Ryd, n.d.)              & 0 & 6.517 & 6.536 &   4.7 & 6.654 &  2.3 & 6.654 &  2.3 & 6.444
\\ \hline						
    \multirow{7}{*}{(\textbf{8})}&
B1 (Val, $n$-$\pi^*$)       & 0 & 4.016 & 4.034 &   7.7 & 4.204 &  8.0 & 4.204 &   8.0 & 3.800\\
&A1 (Val, $\pi$-$\pi^*$)    & + & 4.428 & 4.451 &  21.8 & 4.616 & 21.3 & 4.477 &  20.4 & 4.249\\
&B2 (wCT, $\pi$-$\pi^*$)    &+++& 4.906 & 4.944 &  54.3 & 5.199 & 54.5 & 5.190 &  54.2 & 4.647\\
&A2 (Val, $n$-$\pi^*$)      & + & 5.383 & 5.371 & -17.9 & 5.480 &-18.3 & 5.480 & -18.3 & 5.100\\
&A2 (Val, $n$-$\pi^*$)      & 0 & 5.811 & 5.846 &  -7.6 & 6.012 & -6.4 & 6.012 &  -6.4 & 5.389\\
&A1 (Val, $\pi$-$\pi^*$)    & ++& 5.912 & 5.887 &  27.5 & 6.044 & 28.1 & 5.812 &  26.0 & 5.664\\
&B2 (Val, $\pi$-$\pi^*$)    & + & 6.487 & 6.498 &  12.4 & 6.661 & 12.1 & 6.644 &  12.1 & 6.297
\\ \hline
\multicolumn{2}{c}{ME} &&
    0.217   &   0.216   & -- &   0.365   & -- &   0.306   & --   & --
    \\
\multicolumn{2}{c}{MAE} &&
    0.221   &   0.216   & -- &   0.365   & -- &   0.307   & --   & --
    \\
\multicolumn{2}{c}{SD} &&
    0.140   &   0.130   & -- &   0.157   & -- &   0.180   & --   & --
    \end{tabular}
    }
    \caption{Comparison of vertical singlet–singlet excitation energies (in eV) for charge-transfer states calculated with various EOM-CC/aug-cc-pVTZ models.
    'nc' means the value could not be computed. 
    The dCT {type} and dCT[\%] columns indicate the strength and percentage contribution of the directed charge transfer, respectively.
    TBE stands for the theoretical best estimate.}
    \label{tab:ee-eom}
\end{table}

\subsection*{Doubly Excited States}
To assess the performance of our EOM-CC models for doubly excited states, we selected a subset of the QUEST database consisting of molecules shown in Figure~\ref{fig:structures-double-excitations}.
All selected systems possess at least one low-lying singlet excited state with a significant double excitation character, for which highly accurate theoretical best estimates (TBEs) are available.~\cite{quest-jctc-2025}
These states are labeled as 'dou'/'double' or 'par dou' in the database, indicating doubly excited and pair-doubly excited character, respectively.
In the following subsection, we focus solely on the lowest-lying of these challenging doubly excited states, leaving the rest of the spectrum aside.
Furthermore, all EOM-CC calculations are performed with HF orbitals, as the identification of all targeted doubly-excited states is rather difficult in the localized pCCD-optimized molecular-orbital basis.

The vertical singlet–singlet excitation energies obtained with our EOM-pCCD-based models are presented in Table~\ref{tab:ee-eom-double-excitation}.
For comparison, results from the standard EOM-CCSD method are also included.
Additionally, the table reports the percentage contribution of singly excited configurations (\ce{\%C1}) for all methods.
The value of \ce{\%C1} = 1 corresponds to a purely singly excited transition, \ce{\%C1} = 0 indicates a purely doubly excited transition, and any intermediate value reflects a mixed character of the excitation.
It is important to note that not all doubly excited states can be treated with every EOM model.
These cases are indicated by 'nc' in Table~\ref{tab:ee-eom-double-excitation}.

As shown in Table~\ref{tab:ee-eom-double-excitation}, the EOM-ptCCSD model yields excitation energies and \ce{\%C1} contributions that are nearly identical to those obtained with the standard EOM-CCSD.
The differences are typically very small, usually below 0.1 eV.
Moreover, in all cases where a given doubly excited state could not be computed with EOM-CCSD, the EOM-ptCCSD calculation also failed.
At the same time, the simple EOM-pCCD+S model could compute some of them; exceptions are molecules (\textbf{9}), (\textbf{10}), (\textbf{16}), (\textbf{17}){,} and (\textbf{20}).
However, as observed in other works, the excitation energies computed with EOM-pCCD+S are usually too high and overshoot the reference TBE values by a few eV.~\cite{eom-pccd-jcp-2016, pccd-delaram-rsc-adv-2023, pccd-mocco-galynska-pccp-2024, eom-pccd+s-ct-analysis-jctc-2025,eom-pccd+s-ct-analysis-carbazole-ram-jpca-2025}

Much better results are obtained from the EOM-fpLCCSD and EOM-fpCCSD schemes.
All of them can compute double-excited states for all investigated systems, expect the first excited state of (\textbf{9}) and (\textbf{16}).  
In most cases, both methods yield results quite comparable with the TBE reference, with EOM-fpCCSD being usually the closest. 
The most difficult to describe for these models are the $\Delta_g$ and $\Sigma_g^+$ electronic states of the carbon dimer (\textbf{12}) and the $A_g$ state of glyoxal (\textbf{18}). 
These states are known to have multi-determinant character and, by definition, should not be modeled with a single reference CC method. 
Notably, the doubly excited states of (\textbf{12}) become much more reliable (1.787 eV and 1.917 eV) when the pCCD orbitals are utilized, significantly reducing the error to TBE (2.091 eV and 2.420 eV).  
When we exclude these molecules from our analysis, we find that the EOM-fpLCCSD and EOM-fpCCSD models reproduce the reference TBE data, usually within 0.2-0.5 eV. 
At the same time, the SD drops from 0.799 to 0.277 and from 0.790 to 0.321 for EOM-fpCCSD and EOM-fpLCCSD, respectively. 
In all cases, \ce{\%C1} is almost the same for both models, but quite often different from EOM-ptCCSD and EOM-CCSD.  
Finally, we observe that the \ce{\%C1} value does not correlate with the accuracy of the investigated EOM method.

In summary, the newly introduced EOM-fpCCSD model significantly improves the description of doubly excited states compared to the EOM-CCSD and EOM-ptCCSD models for all investigated systems reported in Table~\ref{tab:ee-eom-double-excitation}.  


\begin{table}[tb]
    \centering
    \resizebox{1\textwidth}{!}{
    \begin{tabular}{c l | cc | cc | cc | cc | cc| c }
    \hline
    Molecule& State 
                    & \multicolumn{2}{c|}{EOM-pCCD+S(HF)} 
                    & \multicolumn{2}{c|}{EOM-fpLCCSD(HF)}
                    & \multicolumn{2}{c|}{EOM-ptCCSD(HF)}
                    & \multicolumn{2}{c|}{EOM-fpCCSD(HF)}
                    & \multicolumn{2}{c|}{EOM-CCSD(HF)}
                    & \multicolumn{1}{c}{TBE Extrapol.~\cite{quest-jctc-2025}} 
                    \\
                    & &
                    \%\ce{C1}&$E_{\rm ex}$ [eV]&
                    \%\ce{C1}&$E_{\rm ex}$ [eV]&
                    \%\ce{C1}&$E_{\rm ex}$ [eV]&
                    \%\ce{C1}&$E_{\rm ex}$ [eV]&
                    \%\ce{C1}&$E_{\rm ex}$ [eV]& $E_{\rm ex}$ [eV]
                    \\
    \hline
  \multirow{2}{*}{(\textbf{9})}&
Ag (Val, double, n,n-$\pi^*$,$\pi^*$)  
& nc & nc
& nc & nc
&nc   & nc
& nc& nc
& nc & nc
& 4.566
\\
&Ag (Val, par double, $\pi$,$\pi-\pi^*$,$\pi^*$)  
& 0.37&  7.961
& 0.20&  6.562
& nc & nc
& 0.23&  6.377
& nc& nc
& 6.351
\\ \hline
  \multirow{2}{*}{(\textbf{10})}&
A1 (Val, double, $\pi$,$\pi-\pi^*$,$\pi^*$)  
& nc &  nc
& 0.05 & 5.375
& nc & nc
& 0.06 & 5.289
&  nc &  nc
& 4.708
\\
&A1 (Val, double, $\pi$,$\pi-\pi^*$,$\pi^*$)
& 0.43& 7.229
& 0.73& 7.041
& 0.88& 6.593
& 0.70 &  6.811
&  0.88 & 6.609
& 6.484 \\ \hline
\multirow{1}{*}{(\textbf{11})}&
Ag (Val, par dou, $\pi-\pi^*$)
& 0.99 & 7.426
& 0.71 & 7.076
&  0.88   & 7.100
& 0.71& 6.851
& 0.88 & 7.123
& 6.515
\\ \hline
  \multirow{2}{*}{(\textbf{12})}&
$\Delta_g$ (Val, dou, $\pi$,$\pi-\sigma$,$\sigma$) 
& 0.00 & 1.608
& 0.00 & 0.314
&  0.00   & 4.934
&0.00 & 0.287
& 0.00 & 4.617
& 2.091
\\
&
$\Sigma_g^+$ (Val, dou, $\pi$,$\pi-\sigma$,$\sigma$)
& 0.00&  1.803
& 0.00& 0.352
& 0.00& 4.947
& 0.00& 0.319
&0.00& 4.763
& 2.420
\\ \hline
  \multirow{2}{*}{(\textbf{13})}&
$\Delta_g$ (Val, dou, n,n-$\pi^*$,$\pi^*$)
& 0.00 & 7.114
& 0.00 & 5.615
& 0.00  & 9.120
& 0.00& 5.577
& 0.00 &   9.209
& 5.230
\\
&
$\Sigma_g^+$ (Val, dou, n,n-$\pi^*$,$\pi^*$) 
& 0.00&  7.520
& 0.00& 6.022
&0.00 & 9.256
& 0.00& 5.984
&0.03 &  9.524 
& 5.908
\\ \hline
  \multirow{1}{*}{(\textbf{14})}&
Ag (Val, dou, $\pi$,$\pi-\pi^*$,$\pi^*$)
& 0.07 & 7.224
&  0.00& 4.353
& 0.75  & 7.381
& 0.00 & 4.272
& 0.74 & 7.400
& 4.036
\\ \hline
\multirow{3}{*}{(\textbf{15})}&
B1 (Val, dou, n,$\pi-\pi^*$,$\pi^*$)
& 1.00 & 3.585
& 0.92 & 3.220
&  0.92 & 2.975
&0.92 & 2.975
&  0.92 & 2.924
& 3.156
\\
& A1 (Val, par dou, $\pi$,$\pi-\pi^*$,$\pi^*$) 
&  0.80&  5.246
& 0.62& 5.462
& 0.91& 5.188
& 0.69&  4.986
& 0.91& 5.114
& 5.329
\\ 
& A1 (Val, dou, n, n-$\pi^*$,$\pi^*$) 
&0.56 & 7.703
&0.19&  5.813
&0.93 &  5.937
&0.23& 5.702
&0.93& 5.894
&5.555
\\ 
\hline
\multirow{3}{*}{(\textbf{16})}&
B1 (Val, dou, n, $\pi-\pi^*$, $\pi^*$)
& nc & nc
& nc  &   nc 
&  nc  &  nc
& nc & nc
& nc & nc
& 5.009
\\
&A1 (Val, par dou, $\pi$, $\pi-\pi^*$, $\pi^*$)
& 0.61&  7.490
&   0.13&  5.891
& nc& nc
& 0.17& 5.778
&nc & nc
& 5.795
\\ 
&A1 (Val, dou, n, n-$\pi^*$, $\pi^*$)
& 0.41 &8.001
& 0.66 & 7.311
& nc &  nc
& 0.64 & 6.986
& nc & nc 
& 6.714
\\ \hline
\multirow{1}{*}{(\textbf{17})}&
Ag (Val, dou, $\pi$,$\pi-\pi^*$,$\pi^*$)
&  nc& nc
&  0.51 & 13.755
&   nc  & nc
& 0.54&   13.663
&nc  & nc
& 12.899
\\ \hline
\multirow{1}{*}{(\textbf{18})}&
Ag (Val, dou, n,n-$\pi^*$,$\pi*$)
&  0.01 & 9.282
&  0.00 &  7.124
&  nc & nc
& 0.00 &  7.074
&nc  & nc
& 5.492
\\ \hline
\multirow{1}{*}{(\textbf{19})}&
Ag (Val, par dou, $\pi-\pi^*$)
&   0.80 & 7.733
& 0.44  & 6.155
&  0.83 & 6.540
& 0.47 &   5.934
&0.83 & 6.571
&5.435
\\ \hline
\multirow{1}{*}{(\textbf{20})}&
A' (Val, dou, n,n-$\pi^*$,$\pi^*$)
& nc & nc
& 0.00  & 5.215
&  0.08   &  9.604
& 0.0&  5.141
& 0.08 & 9.633
& 4.732
\\ \hline
\multirow{1}{*}{(\textbf{21})}&
A' (Val, dou, n,n-$\pi^*$,$\pi^*$)
& 0.03 & 11.228
& 0.03 & 8.229
&  nc&  nc
& 0.03 &  8.113
& nc & nc
& 7.969
\\ \hline
\multirow{1}{*}{(\textbf{22})}&
A' (Val, dou, n,n-$\pi^*$,$\pi^*$)
& 0.00 & 8.504
& 0.00& 4.859
& 0.04 & 8.880
& 0.0 & 4.804
&  0.04 &8.863
& 4.333
\\ \hline
\multirow{1}{*}{(\textbf{23})}&
Ag (Val, dou, n,n-$\pi^*$,$\pi^*$)
& 0.00 & 7.596
& 0.00&   6.036
& nc & nc
& 0.00  &  5.998
&nc & nc 
& 4.951
\\
\hline
\multicolumn{2}{c}{ME} &&
1.694 & &  0.270 & &  1.943 & &  0.134 & &  1.928\\
\multicolumn{2}{c}{MAE} &&
1.825 &  & 0.637 &  & 1.993 &  & 0.558 &  & 1.996\\
\multicolumn{2}{c}{SD} &&
1.359 & &  0.799 & &  1.799 & &  0.790 & &  1.823\\
\hline
\multicolumn{2}{c}{\phantom{A}ME w/o (\textbf{12}) and (\textbf{18})} &&
1.853 & &  0.438 & &  1.943 & &  0.286 & &  1.928\\
\multicolumn{2}{c}{MAE w/o (\textbf{12}) and (\textbf{18})} &&
1.864 &  & 0.438 &  & 1.993 &  & 0.346 &  & 1.996\\
\multicolumn{2}{c}{\phantom{M}SD w/o (\textbf{12}) and (\textbf{18})} &&
1.108 & &  0.277 & &  1.799 & &  0.321 & &  1.823
\\
\hline \hline
    \end{tabular}
}
    \caption{Vertical singlet–singlet excitation energies (in eV) for states with significant doubly excited character calculated using various EOM-CC/aug-cc-pVTZ models.
    `nc' denotes values that could not be computed.
    TBE stands for the theoretical best estimate.
    {\mbox{\%\ce{C1}} is the percentage contribution of singly excited configurations.}
    }
    \label{tab:ee-eom-double-excitation}
\end{table}

\section*{Conclusions}
In this work, we introduced and benchmarked a new EOM-CC method based on a pCCD reference: the frozen-pair EOM-CCSD (EOM-fpCCSD) model.
We have shown that EOM-fpCCSD reliably describes low-lying singlet–singlet vertical excitation energies across a wide range of molecular systems and excitation characters.
For CT states from the QUEST database, EOM-fpCCSD yields excitation energies closest to those obtained with standard EOM-CCSD, outperforming EOM-ptCCSD when using pCCD natural orbitals.
The choice of orbital basis has only a minor effect on the computed excitation energies (differences between EOM-CCSD(HF) and EOM-CCSD(pCCD) are typically below 0.01 eV), while enabling analysis of the directed CT character.
The strength of the directed CT, quantified by dCT and dCT[\%], is found to be very similar across all three methods.
Most importantly, the DAISpY-based domain-based CT analysis allows us to quantify the domain-based charge flow for each excited state, facilitating a unique classification of CT states.

The true advantage of EOM-fpCCSD becomes evident for states with significant doubly excited character.
While EOM-ptCCSD largely inherits the well-known limitations of EOM-CCSD for these challenging excitations, EOM-fpCCSD substantially improves both the accuracy and the stability of the calculations.
In most cases, the errors relative to the theoretical best estimates are reduced from several eVs (EOM-CCSD and EOM-ptCCSD) to 0.2–0.5 eV.
Moreover, EOM-fpCCSD converges for several doubly excited states that cannot be described by the standard EOM-CCSD or EOM-ptCCSD approaches.
Additionally, EOM-fpCCSD slightly outperforms its linearized EOM-fpLCCSD variant. 
Thus, decoupling the seniority-zero from other seniority sectors helps to improve the description of challenging excited states.
Nevertheless, EOM-fpCCSD still struggles with states that exhibit strong multi-reference character, indicating that further development is required for a fully satisfactory description of highly correlated excited states.
{Finally, to further assess the potential of the proposed EOM-fpCCSD workflow, we need to investigate cases beyond the Franck-Condon region.
Possible examples include monitoring the transition from locally-excited to charge-transfer states during the rotation or twist of the dimethylamino group in dimethylaminobenzonitrile,\mbox{~\cite{dimethyl-amino-benzonitrile-twist-jacs-2004, reference-ict-jctc-2021}} that is, going from (\textbf{3p}) to (\textbf{3t}), or the evolution of doubly-excited states along potential energy surfaces (such as the automerization of cyclobutadiene).\mbox{~\cite{cyclobutadiene_automerization_ex_state_benchmark_loos_jpca_2022}
}}

In summary, the proposed EOM-fpCCSD model offers a computationally efficient ($\mathcal{O}(N^6)$) alternative to conventional EOM-CCSD.
It matches the accuracy of EOM-CCSD for singly excited states while providing a significant improvement for single-reference-dominated doubly excited states.
These results highlight the promising potential of pCCD-based EOM methods for reliably describing complex electronic excitations, including those encountered in organic electronic materials.

\section*{Data availability}
The data supporting this article have been included as part of the Supplementary Information.
The PyBEST code is available on Zenodo at \url{https://zenodo.org/records/10069179} and on PyPI at \url{https://pypi.org/project/pybest/}.
The version of the code employed for this study is version v2.2.0.dev0.

\section*{Acknowledgements}
Funded/Co-funded by the European Union (ERC, DRESSED-pCCD, 101077420).
Views and opinions expressed are, however, those of the author(s) only and do not necessarily reflect those of the European Union or the European Research Council. Neither the European Union nor the granting authority can be held responsible for them. 
P.T.~acknowledge financial support from the SONATA BIS research grant from the National Science Centre, Poland (Grant No. 2021/42/E/ST4/00302).
This work was completed in part at the Poland Open Hackathon, part of the Open Hackathons program. 
The authors acknowledge OpenACC-Standard.org for their support.
We gratefully acknowledge Polish high-performance computing infrastructure PLGrid (HPC Centers: ACK Cyfronet AGH and WCSS) for providing computer facilities and support within computational grant no. PLG/2025/018840.

\begin{NoHyper}
\footnotetext{\textit{$^{a}$~Institute of Physics, Faculty of Physics, Astronomy, and Informatics, Nicolaus Copernicus University in Toruń, Grudziadzka 5, 87-100 Toruń, Poland. E-mail: k.boguslawski@umk.pl}}
\footnotetext{\textit{$^{b}$~Institute of Physics, Faculty of Physics, Astronomy, and Informatics, Nicolaus Copernicus University in Toruń, Grudziadzka 5, 87-100 Toruń, Poland.}}

\footnotetext{\dag~Supplementary Information available: {Working equations for EOM-fpCCSD and EOM-ptCCSD. Raw data containing all excitation energies and CT matrix elements for different basis sets.} See DOI: 00.0000/00000000.}
\end{NoHyper}



\section*{Author Contributions}
K.B.: Conceptualization (equal), Data curation (equal), Formal analysis (equal), Funding acquisition (lead), Investigation (equal), Methodology (lead), Project administration (lead), Software (lead), Validation (equal), Visualization (lead), and Writing – original draft (equal).

P.T.: Conceptualization (equal), Data curation (equal), Formal analysis (equal), Investigation (equal), Resources (lead), Validation (equal), Visualization (equal), and Writing – original draft (equal).

\section*{Conflicts of interest}
There are no conflicts to declare.






\bibliographystyle{rsc} 
\bibliography{rsc} 

\end{document}